\documentstyle[10pt]{article}
\begin{document}
\hfill{UM-P-95/63}

\hfill{RCEHP-95/17}

\vspace{1cm}

\centerline{\bf \large NEW ISSUES IN QUARK-LEPTON SYMMETRY\footnote{
Invited talk given at Valencia 95, to appear in the Proceedings.}}

\vspace{5mm}

\centerline{\large R.R. VOLKAS }

\vspace{5mm}

\centerline{\it Research Centre for High Energy Physics, School of Physics}
\centerline{\it University of Melbourne, Parkville 3052, Australia}

\vspace{1cm}

\centerline{Abstract}

Two models incorporating different forms of
spontaneously broken quark-lepton symmetry are discussed. Both
models are constructed so that quark-lepton symmetry can be broken at
as low an energy scale as phenomenology allows, thus maximising their
testability. The first model uses the Pati-Salam gauge group
SU(4)$\otimes$SU(2)$_L\otimes$SU(2)$_R$. The main analysis presented
here concerns large threshold corrections to the $m_b = m_{\tau}$
tree-level mass relation. The second model is a new and
simplified theory using discrete quark-lepton symmetry which has a
very economical Higgs sector and elegantly avoids a potential
quark-lepton mass relation problem.

\section{Introduction}

The idea that quarks and leptons are related in a pairwise fashion has
a long history. In the late 1950's Marshak discussed an evident
``quark-hadron correspondence'' whereby the leptons $e$, $\nu_e$ and
$\mu$ were associated with the hadrons $p$, $n$ and $\Lambda$. This
turned into a ``quark-lepton correspondence'' between $e$, $\nu_e$,
$\mu$ and $d$, $u$, $s$ once quark substructure was revealed. With
the discovery of $\nu_{\mu}$, the quark-lepton correspondence idea was
used to postulate the existence of a fourth quark $c$. This remarkable
prediction bore fruit with the discovery of charm in the mid 1970's.
The correspondence between quarks and leptons was subsequently found
to also hold for third generation fermions.

The pairwise association between quarks and leptons is so striking
that it makes sense to suppose that the fundamental Lagrangian of the
world actually treats quarks and leptons on an equal footing, with
their disparate properties ascribed to a spontaneous violation of the
exact quark-lepton symmetry. Such a symmetry can either be continuous
or discrete. Candidates for a continuous symmetry of this nature are
Pati-Salam SU(4) \cite{ps} and the various grand unification groups.
Alternatively, a discrete symmetry between quarks and leptons can be
introduced by postulating a spontaneously broken colour SU(3) group
for leptons \cite{fl}.

In this talk I will describe two pieces of recent work in this area.
The first piece of work analyses some of the implications of having
Pati-Salam SU(4) broken at the lowest energy scale that phenomenology
allows \cite{willenbrock} (1000 TeV).
I will discuss how best to treat the neutrino mass
problem, followed by the core of the analysis which is a detailed
computation of radiative corrections to the tree-level mass relation
$m_b = m_{\tau}$. I will show here that threshold corrections due to
charged Higgs boson graphs can be very large. The second piece of work
is a reconstruction of the quark-lepton discrete symmetry idea in such
a way that the Higgs sector of the model is kept as simple as possible.
I will also demonstrate that the resulting theory deals with the
quark-lepton mass relation ``problem'' in an elegant way.

\section{Low-scale Pati-Salam SU(4)}

The following discussion is based on Ref.\cite{volkas}.
Under the Pati-Salam gauge group $G_{\rm PS} =
$SU(4)$\otimes$SU(2)$_L\otimes$SU(2)$_R$, a fermion generation appears
as follows:
\begin{equation}
f_L \sim (4,2,1),\qquad f_R \sim (4,1,2).
\end{equation}
The SU(4) factor transforms quarks into leptons in a continuous way.
It contains the subgroup SU(3)$_c\otimes$U(1)$_{B-L}$. The linear
combination $2I_R + (B-L)$, where $I_R$ is the diagonal generator
of SU(2)$_R$, is identified as weak hypercharge $Y$. I do not impose a
discrete left-right symmetry on the model.

The model has a two stage symmetry breaking chain. At a relatively
high scale $M_{\rm PS}$ SU(4) and SU(2)$_R$ are broken simultaneously,
which is followed by electroweak symmetry breaking at its usual scale.
Let us discuss electroweak symmetry breaking first. This is achieved
by introducing a complex bidoublet Higgs field $\Phi \sim (1,2,2)$.
The Yukawa Lagrangian
\begin{equation}
{\cal L}_{\rm Yuk} = \lambda_1 {\rm Tr}[\overline{f}_L \Phi f_R] +
\lambda_2 {\rm Tr}[\overline{f}_L \Phi^c f_R] + {\rm H.c.}
\end{equation}
where $\Phi^c \equiv \tau_2 \Phi^* \tau_2$
yields the tree-level mass relations
\begin{equation}
m_b = m_{\tau}\quad {\rm and}\quad m_t = m^{\rm Dirac}_{\nu_3}
\end{equation}
given a nonzero vacuum expectation value for $\Phi$ (the
fermion multiplets are written as $2 \times 4$ matrices in the
above). Similar relations
hold for the first two generations, but the present analysis will
confine itself to the third generation. The deep problem of
generation structure will not be tackled here.

What is one to do with these mass relations? The equality between
$m_t$ and the neutrino Dirac mass is usually taken care of through the
see-saw mechanism \cite{seesaw} by using a $\Delta \sim (10,1,3)$
Higgs multiplet
to generate a large Majorana mass for the right-handed tau neutrino.
This also breaks SU(4)$\otimes$SU(2)$_R$ down to
SU(3)$_c\otimes$U(1)$_Y$. However, the cosmological closure bound on
the light mass eigenstate $m_t^2/\langle\Delta\rangle < 30$ eV leads
to $\langle\Delta\rangle \sim 10^{12}$ TeV. This is much higher than
the phenomenological lower bound of 1000 TeV, and thus use of the
standard see-saw mechanism would undermine the motivation for the
analysis.

An interesting alternative is to use what I will call the ``$3 \times 3$
see-saw mechanism'' \cite{wyler}.
One introduces a singlet fermion $S_L \sim
(1,1,1)$ and the simple Higgs multiplet $\chi \sim (4,1,2)$. The
non-electroweak Yukawa Lagrangian
\begin{equation}
{\cal L}_{\rm Yuk} = n \overline{S}_L
{\rm Tr} [\chi^{\dagger} f_R] + {\rm H.c.}
\end{equation}
when combined with the electroweak Yukawa terms above yield a neutrino
mass matrix of the form
\begin{equation}
\left( \begin{array}{ccc}
0\ & m_t\ & 0\ \\ m_t\ & 0\ & n\langle\chi\rangle\ \\ 0\ &
n\langle\chi\rangle\ & 0\
\end{array} \right)
\end{equation}
in the $[\nu_L, (\nu_R)^c, S_L]$ basis. This produces one massless
eigenstate which we identify with the standard neutrino, and a massive
Dirac neutrino. For $n\langle\chi\rangle \gg m_t$, the massless state
has approximately standard electroweak interactions. Because the light
eigenstate is massless for all values of the nonzero entries of the
mass matrix, $\langle\chi\rangle$ can be reduced to about 1000 TeV.

The $m_b = m_{\tau}$ relation receives radiative corrections of two
types: those associated with large logarithms which are best
calculated through renormalisation group equations, and threshold
corrections which depend on either heavy/heavy or light/light mass
ratios. It is interesting that renormalisation group evolution using
standard particles in the loops merges $m_b$ with $m_{\tau}$ at about
1000 TeV \cite{arason}! Therefore if Pati-Salam SU(4) is important for
the fermion mass problem, an indirect indication through $K^0 \to
e^{\pm}\mu^{\mp}$ may be just around the corner. But this renormalisation
group result is only relevant if threshold corrections are not too
large. It turns out that there is a generically large high-scale
threshold correction due to loops containing the physical charged
Higgs boson. See Ref.\cite{volkas} for details. These graphs produce a
threshold correction given by
\begin{equation}
m_{\tau}-m_b \simeq -
\frac{1}{16\pi^2}\frac{m_s^2-m_t^2}{m_H^2-m_s^2}
\frac{m_t(m_t-m\sin2\omega)(m_t\sin2\omega-m)}{(u_1^2 + u_2^2)\cos^2
2\omega}\ln\frac{m_s^2}{m_H^2},
\end{equation}
where $m_s$ is the heavy Dirac neutrino mass, $m_H$ is the charged
Higgs boson mass, $m$ is the ``common'' tree-level
mass for $b$ and $\tau$, $u_{1,2}$ are
the electroweak breaking VEVs and $\tan\omega = u_2/u_1$.

This threshold correction can clearly produce a mass difference
between $m_{\tau}$ and $m_b$ of the order of a GeV, provided an
accidental cancellation between $m$ and $m_t\sin2\omega$ does not
occur. The ``common'' mass $m$ of $\tau$ and $b$ at $M_{\rm PS}$
must be about the same as the measured $m_{\tau}$, namely 1.8 GeV. The
above threshold correction can therefore alter the initial ratio
$m_b/m_{\tau}$ by up to $50\%$. This correction is thus as numerically
significant as those incorporated through the renormalisation group.
The sign of the correction depends on the unknown parameter $\omega$
and therefore cannot be predicted. It can either raise or lower the
mass ratio by up to $50\%$.

This calculation demonstrates that generally speaking one must take
care in the use of renormalisation group evolution to predict
low-energy masses.

\section{New and improved quark-lepton symmetric model}

The idea of discrete quark-lepton symmetry was introduced in
Ref.\cite{fl}. It is based on the gauge group
\begin{equation}
G_{q\ell} = {\rm SU}(3)_{\ell} \otimes {\rm SU}(3)_q \otimes
{\rm SU}(2)_L \otimes {\rm U}(1)_X,
\end{equation}
where SU(3)$_{\ell}$ is leptonic colour, SU(3)$_q$ is ordinary colour
and $X$ is an Abelian charge different from $Y$. A generation of
fermions is placed into the following multiplet pattern:
\begin{eqnarray}
& Q_L \sim (1,3,2)(1/3),\qquad u_R \sim (1,3,1)(4/3),\qquad d_R \sim
(1,3,1)(-2/3), &\ \nonumber\\
& F_L \sim (3,1,2)(-1/3),\qquad E_R \sim (3,1,1)(-4/3),\qquad N_R \sim
(3,1,1)(2/3). &\
\end{eqnarray}
The standard leptons lie within $F_L$, $E_R$ and $N_R$ in a manner
specified by the way $G_{q\ell}$ is spontaneously broken. The
multiplet assignment above allows us to define a discrete quark-lepton
symmetry by
\begin{eqnarray}
& Q_L \leftrightarrow F_L,\qquad u_R \leftrightarrow E_R,\qquad d_R
\leftrightarrow N_R, &\ \nonumber\\
& G^{\mu}_q \leftrightarrow G^{\mu}_{\ell}\qquad {\rm and}\qquad
C^{\mu} \leftrightarrow -C^{\mu}, &\
\end{eqnarray}
where $G^{\mu}_{q,\ell}$ are the gauge bosons of SU(3)$_{q,\ell}$ and
$C^{\mu}$ is the gauge boson of U(1)$_X$.

The original quark-lepton symmetric models
were based on $G_{q\ell}$ breaking achieved through the Higgs
multiplet $\chi \sim (3,1,1)(2/3)$ that appears in the Yukawa
Lagrangian
\begin{equation}
{\cal L}_{\rm Yuk} = h_1 \overline{(F_L)^c} F_L \chi
+ h_2 \overline{(E_R)^c} N_R \chi + {\rm H.c.}
\label{nonewyuk}
\end{equation}
(A quark-lepton symmetric partner for $\chi$ is also introduced with a
corresponding Yukawa Lagrangian.)
A nonzero VEV for $\chi$ breaks $G_{q\ell}$ down to
SU(2)$'\otimes G_{SM}$, where SU(2)$'$ is a subgroup of leptonic
colour that is necessarily unbroken. The weak hypercharge is
identified as $Y = X + T_8/3$,
where $T_8 = {\rm diag}(-2,1,1)$ is a generator of leptonic colour.
The standard leptons are then the $T_8 = -2$ components, while the
$T_8 = 1$ states are exotic fermions.

The minimal electroweak Higgs sector consists of a single doublet
$\phi \sim (1,1,2)(1)$ as in the Standard Model. Under discrete
quark-lepton symmetry it transforms into its charge conjugate field
$\phi^c = i\tau_2 \phi^*$. The electroweak Yukawa Lagrangian
\begin{equation}
{\cal L}_{\rm Yuk} = \lambda_1 (\bar{Q}_L u_R \phi^c + \bar{F}_L E_R
\phi) + \lambda_2 (\bar{Q}_L d_R \phi + \bar{F}_L N_R \phi^c) + {\rm
H.c.}
\label{ewyuk}
\end{equation}
then yields the tree-level mass relations $m_u = m_e$ and $m_d =
m^{\rm Dirac}_{\nu}$.

The second of these relations can be made phenomenologically
acceptable by using the standard see-saw mechanism through the
analogue of the Higgs field $\Delta$ of the Pati-Salam model discussed
previously. In this case, $\Delta \sim (\bar{6},1,1)(-4/3)$. The $m_u
= m_e$ relation can only be dealt with by adding a second electroweak
Higgs doublet which allows one to completely evade mass relations
between quarks and leptons.

A Higgs sector consisting of two $\phi$'s, a $\chi$ and a $\Delta$ is
rather cumbersome. It turns out that there is a much simpler way of
constructing the model, which I now describe. This work is more fully
explained in Ref.\cite{fv} which in turn is based in part on
Ref.\cite{volkas2}. The new Higgs sector consists of only one $\phi$
and two $\chi$-like fields I will call $\chi$ and $\xi$ (plus their
discrete symmetry partners). There is no
$\Delta$ field because I am again going to use singlet fermions $S_L$
together with an analogue of the $3 \times 3$ see-saw mechanism
employed in the previous section. The VEV pattern for $\chi$ and $\xi$
is the most general one possible, namely
$\langle\chi\rangle = (v,0,0)^T$ and $\langle\xi\rangle =
(w_1,w_2,0)^T$ which completely breaks leptonic colour SU(3)$_{\ell}$.
There is no unbroken SU(2)$'$ subgroup, and all components of the
leptonic triplets become integrally charged
[$Y = X + T_8/3 - T_3$ here
where $T_3 = {\rm diag}(0,1,-1)$]. The standard leptons are then a
linear combination of more than one leptonic colour.

The Yukawa Lagrangian of the model contains the electroweak Yukawa
terms of Eq.~\ref{ewyuk}, the non-electroweak terms of
Eq.~\ref{nonewyuk} extended to include $\xi$ in an obvious way,
together with mixing terms between the $S_L$'s and $N_R$'s given by
\begin{equation}
{\cal L}_{\rm Yuk} = n_1 \overline{S}_L \chi^{\dagger} N_R + n_2
\overline{S}_L \xi^{\dagger} N_R + {\rm H.c.}
\end{equation}
The neutrino mass matrix consists of a $7 \times 7$ version of the $3
\times 3$ see-saw mechanism used in my Pati-Salam model. I will not
display it here, except to say that the electroweak breaking entries
in it are specified in terms of the quark masses and that the lightest
mass eigenstate has close to standard electroweak interactions.

The charged lepton mass matrix is given by
\begin{equation}
\left( \begin{array}{ccc}
m_u\ & 0\ & M_3\ \\ 0\ & m_u\ & M_1\ \\ M_4\ & M_2\ & m_d\
\end{array} \right)
\end{equation}
in the basis $(e_1, e_2, e_3)$ where the subscript labels the three
leptonic colours. The $M_i$ are large masses
proportional the the VEVs of $\chi$ and $\xi$.
The smallest mass eigenvalue is given by
$m_e = m_u\cos(\beta_1 - \beta_2) \le m_u$
where $\tan\beta_1 \equiv M_2/M_4$ and $\tan\beta_2 \equiv M_1/M_3$. Note
that the $m_u = m_e$ mass relation does not arise here, and
furthermore that $m_e$ is necessarily less than $m_u$ assuming large
$M_i$ and sufficiently small intergenerational mixing (this type of
result was first found in Ref.\cite{volkas2}). It is straightforward
to check that the fields corresponding to $m_e$ have the correct
electroweak interactions to a good approximation.

This reconstruction of the idea of discrete quark-lepton symmetry
produces a simpler model with regard to the Higgs sector than previous
versions, and it has no quark-lepton mass relation problems. In this
sense the above theory can be regarded as an improvement on earlier
models.

\vspace{1cm}

\centerline{\large Acknowledgments}
This work was supported by the Australian Research Council.

\end{document}